\documentclass[aps,pra,preprint,showpacs,showkeys,nofootinbib]{revtex4-1}
\usepackage[dvips]{graphicx}
\usepackage[normalem]{ulem}
\usepackage{amsmath}
\begin{document}
\title{The role of positronium decoherence in the studies of positron annihilation in matter}
\author{M.~Pietrow}
\email{e-mail: mrk@kft.umcs.lublin.pl; phone: +48 815376261, fax: +48 815376191.}
\affiliation{Institute of Physics, M. Curie-Sk{\l}odowska University, ul. Pl. M. Curie-Sk{\l}odowskiej~1, 20-031 Lublin, Poland}
\author{P.~S{\l}omski}
\affiliation{Geographic Information Systems Development Company \emph{Martinex}, ul. Me{\l}giewska~95, 21-040 {\'S}widnik, Poland}
\begin{abstract}
A small difference between the energy of the para-positronium (p-Ps) and ortho-positronium (o-Ps) states suggests the possibility of superposition of p-Ps and o-Ps during the formation of positronium (Ps) from pre-Ps terminating its migration in the matter in a void ('free volume').\\
It is shown that such a superposition decohere in the basis of p-Ps and o-Ps and the decoherence time is estimated. The time scale of the decoherence estimated here motivates respective correction in decomposition of the positron annihilation lifetime spectra. The way of the correction is sketched. The timescale of the decoherence suggests a need of awareness when experimental data from positron annihilation techniques are processed. More generally, the superposited state of Ps should contribute to the evolution theory of positronium in matter.
\end{abstract}
\maketitle
\section{Introduction}
The most powerful model describing the formation of the positronium atom in the matter was developed by Stepanov and Byakov \cite{Stepanov03}. A positron from the irradiation source (the kinetic energy of positron produced by the most exploited isotope $^{22}$Na \cite{Firestone96} is less than 0.5~MeV) passing through the sample creates a set of electrons, ions, radicals etc. losing its own energy. According to this model, at the beginning, each of the electrons caused by ionization receives about 1~keV of the kinetic energy and is able to ionize other particles. Further ionizations release 30-100 eV each and produce new charged particles which form clusters called \emph{spurs}. If the energy of the positron decreases to about 1 keV, the spacings of ionization spurs become less distant and overlap each other. Finally, the positron ends up in the spreading cloud of ionized particles called \emph{blob} with the energy lower than the ionization level. Further loss of the positron energy is transformed into vibrations of molecules and reaching thermalization energy the production of positronium atom arises as a profitable process.\\
Before the formation of positronium atom, a positron and an electron can leave the blob as a weakly interacting, delocalized system (pre-Ps) migrating to near free volume where it forms the localized Ps. It is usually believed that at this time the whole population of Ps consists of p-Ps (spin $S$=0) and o-Ps ($S$=1) so the spin value for each particle is definite. Such a Ps can annihilate in different ways: 1). the annihilation of p-Ps produces two gamma quanta, 2). o-Ps produces three gamma quanta, 3). each Ps can annihilate with an electron captured from the environment (\emph{pick-off} process). The probability of pick-off annihilation increases with the local electron density and dominates in the case of annihilation in condensed matter. In the case of o-Ps the annihilation via pick-off is a two-quantum process. However, when the free volume is not reached, the positron annihilates as a quasi-free particle with one of the electrons from the bulk (in particular those from the blob).\\
The most serviceable channel of the positron annihilation in a matter is long-living o-Ps which succeeded in reaching the free volume and annihilating by pick-off. In order to separate the long living component from the lifetime spectra, they have to be decomposed into elementary channels of annihilation (free positron annihilation, ortho- and para-Ps annihilation, etc.). The exponential form of annihilation for each channel is commonly assumed.
\section{Motivation}
The papers exploiting the blob model assume the definite value of spin number for pre-Ps (see positronium formation formulas in \cite{He05}). However, as the difference of energies of p-Ps and o-Ps states in the vacuum is about $8.4\cdot 10^{-4}$ eV \cite{Mogensen94}, one can expect that the spin quantity could change during the migration of a pre-Ps through the sample due to the intermolecular interactions. It seems to be an oversimplification considering the definite spin value only, e.g. only o-pre-Ps. Instead, one should suppose that quasi-free Ps enters a free volume as a superposition of para and ortho states and thus the spin number could not be a proper number for such initial state.\\
The value of Ps spin plays an important role in further analysis of Ps life. The annihilation of Ps, as an electromagnetic process, should obey the principle of conservation of the parity (here charge parity). This number for Ps in its ground state with the spin value $S$ is $\pi_C=(-1)^S$ whereas for $n$ photons \uwave{it} is $\pi_C=(-1)^n$ \cite{Greiner96}. So until the spin number is well defined, the system cannot decay with the emission of any definite $n$ photons. Thus supposing that Ps is the superposition of p-Ps and o-Ps after entering free volume, one could expect that Ps can ``live'' in the free volume without giving an identifying decay signal (the number of photons is undefined) until it becomes p-Ps \emph{or} o-Ps. This implies that one cannot decompose the PALS spectra simply into the exponential curves for p-Ps and o-Ps because the decay (two-quantum, for example) cannot start just after the birth of Ps. In our opinion, as long as there exists the population of Ps which is superposition of spins, one should make the correction when decomposing the PALS spectra (the analysis of the correction way is not the subject of this paper but an example of applying the correction is shown in Appendix \ref{sec:appendix1}). In this paper we show that it is expected to decohere Ps in the p-Ps and o-Ps basis and we estimate the time of decoherence.
\section{Rate of decoherence. Method of calculation}
The most convenient hamiltonian to consider the spin system interacting with the environment of spins (\emph{bath}) is the Heisenberg hamiltonian \cite{Zurek05} which is the phenomenological expression incorporating an overlapping of wave functions of electrons of the system and an environment \cite{Auerbach94,Nolting09}. The problem of the decoherence of certain spin systems, in particular consisting of a pair of particles interacting with the bath, was considered in \cite{Dobrovitski03,Dobrovitski04,Dobrovitski04a} which support our calculation method.\\
Let us consider the hamiltonian for interaction of Ps and the molecules from the wall of the free volume
\begin{equation}
H=J{{\hbar^2}\over{4}}\sum_{i=3}^{N+2}\bar{\sigma}^{(2)}\circ\bar{\sigma}^{(i)},
\label{hamiltonian}
\end{equation}
where $J$- coupling constant, $\sigma^{(2)}$- Pauli matrices for the electron in Ps, $\sigma^{(i)}$- Pauli matrices for the $i$-th (of $N$) electron from an environment, the sign $\circ$ denotes a scalar product.\\
We assume the existence of free volumes where Ps could live sufficiently long time\footnote{Our model seems to be the most suitable for the solid state of matter. Probably the model could be applied also to the liquid state where, according to the recent knowledge, the bubbles are formed as free volumes for Ps.}. As the environment we mean atoms which form the wall of the free volume. There is a wide spectrum of materials where such free volumes exist (molecular crystals or polymers). In all these cases we suppose the atoms have certain electron density around (inside) the free volume. These valence electrons can interact with an electron from Ps by exchange interaction and resulting in the pick-off process of Ps.\\
The radii of the free volumes are in the range considered by the PALS method, i.e. from about 1$\AA$ to hundreds of nanometers. The wavelength of valence electrons lies in the range of angstroms. So one can expect in general that Ps in the free volume interacts with the electrons only momentarily. Generally, it is not obvious how many ($N$ variable in this work) electrons interact with Ps at the same time.
\\
Suppose the initial (just after the localization of Ps in the free volume) state of the whole system is $\rho_0\equiv |\Psi_0\rangle\langle\Psi_0|$, where
\begin{equation}
|\Psi_0\rangle=|Ps\rangle_0\prod_{i=3}^{N+2}{|s_i\rangle}\equiv |Ps\rangle_0|S\rangle
\label{stan}
\end{equation}
and
\begin{equation}
|Ps\rangle_0=\frac{1}{2}(|0,0\rangle+|1,-1\rangle+|1,0\rangle+|1,+1\rangle).
\end{equation}
The ket $|S\rangle$ denotes the state of electrons of the environment whereas the numbers in the kets are spin $s$ and its $z$-th projection $s_z$ for Ps, respectively. The latter equation expresses the supposition that after many interactions with the bulk, Ps, just after entering into the free volume, does not distinguish any particular Ps state.\\
The evolution in time gives the following expression for the whole state at the instant $t$
\begin{equation}
\rho(t)=|\Psi_t\rangle\langle\Psi_t|,
\end{equation}
where $|\Psi_t\rangle=e^{-iHt/\hbar}|\Psi_0\rangle$.
For the positronium only, we calculate the state at $t$ as
\begin{equation}
\rho_{Ps}(t)=Tr_{env}[\rho(t)]
\end{equation}
tracing over the environmental space.\\
In order to estimate the admixture of p-Ps in o-Ps at $t$ we calculate some nondiagonal elements of $\rho_{Ps}(t)$. More precisely, we consider the expression
\begin{equation}
S(t)=\sum_{s_z=-1}^{+1} |\langle s=0|\rho_{Ps}(t)|s=1,s_z\rangle|^2
\label{S}
\end{equation}
which is expected to collapse with time. The time when it reaches
the minimum value we define as the time of the decoherence, $T_d$.\\
Because of the mathematical problems with calculating the whole $\rho(t)$ analytically, we made the following simplification
\begin{equation}
|\Psi_t\rangle=e^{-iHt/\hbar}|\Psi_0\rangle\simeq\sum_{j=0}^{n}\frac{1}{j!}(-\frac{itH}{\hbar})^j\ |\Psi_0\rangle,
\label{expansion}
\end{equation}
where the value $n$ is fixed in this formula. This truncation at appropriate $n$ still allows to approximate $T_d$ sufficiently for our aims (see Appendix \ref{sec:appendix2}).
\section{Results and discussion}
\begin{figure}
\centering
\includegraphics[width=0.7\textwidth]{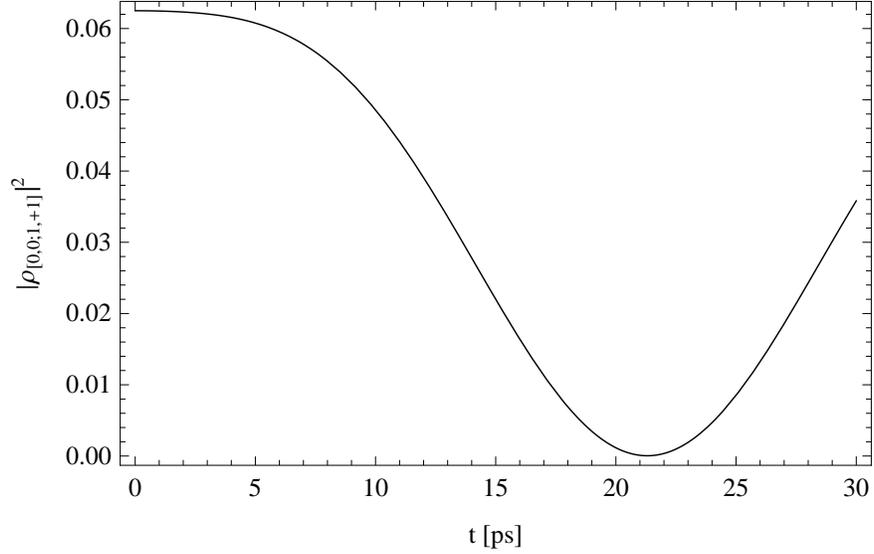}
\caption{$|\rho_{[0,0;1,+1]}(t)|^2$ calculated for a random initial state vector~(\ref{stan}). The interaction of Ps with $N$=2 electrons from the wall was assumed and $n$ in (\ref{expansion}) was set 10. The time to the minimal value of the function is regarded as the decoherence time.}
\label{fig.1}
\end{figure}

\begin{figure}
\centering
\includegraphics[width=0.7\textwidth]{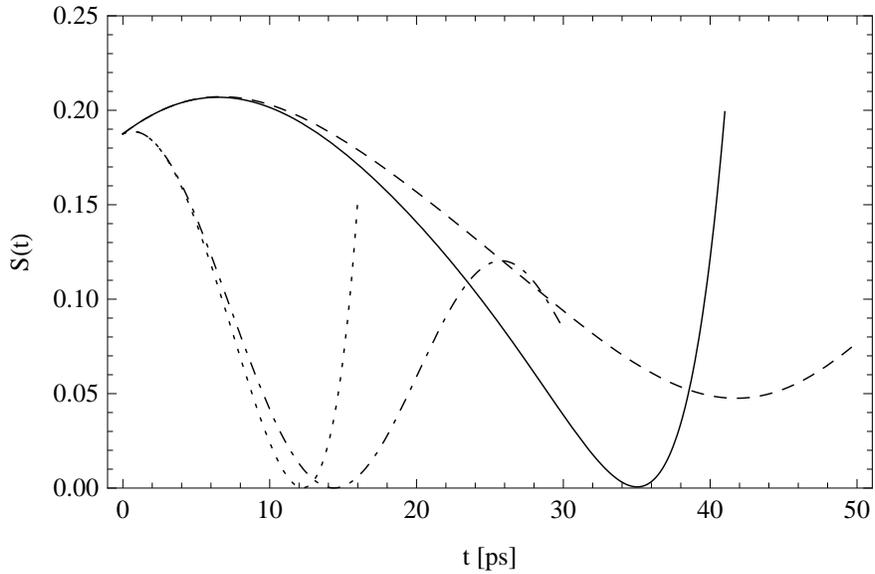}
\caption{$S(t)$ averaged over 100 randomly chosen samples of the initial state vector for different electrons interacting with Ps atom ($N$). The comparison of calculations with truncation of the evolution operator by $n$=3 and by $n$=10.  The results are shown for $N$=1 (solid and dashed line, respectively) and for $N$=4 (dotted and dashdotted line, respectively).}
\label{fig.2}
\end{figure}

\begin{figure}
\centering
\includegraphics[width=0.7\textwidth]{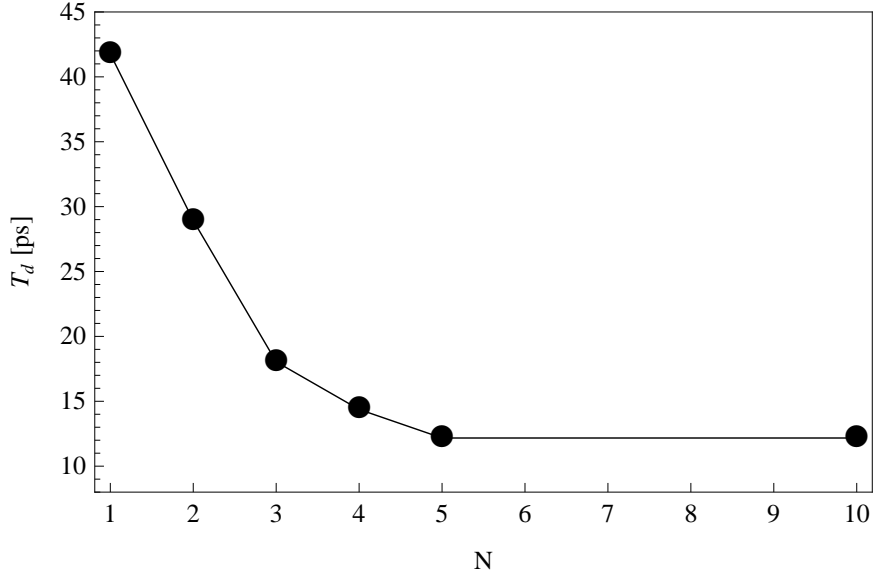}
\caption{The decoherence time $T_d$ as a function of the number $N$ of interacting electrons with Ps. The value is averaged over 20 randomly chosen initial state vectors for $n$=7.}
\label{fig.3}
\end{figure}

\begin{figure}
\centering
\includegraphics[width=0.7\textwidth]{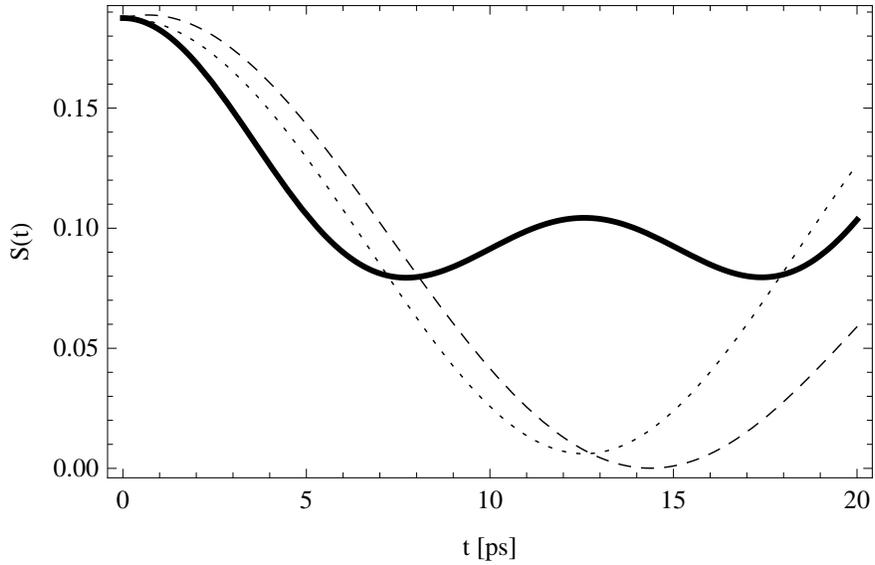}
\caption{The presence of the magnetization of the medium. The curves are calculated for $N$=4, $n$=10. The curve without magnetization (randomly situated spins of the environment) was obtained as an average over 100 random samples of spin set; dashed line. The bold line-- all spins are situated along the $z$-axis, the dotted line- the direction of all environmental spins are situated along the $x$-axis.}
\label{fig2r}
\end{figure}

\begin{figure}
\centering
\includegraphics[width=0.7\textwidth]{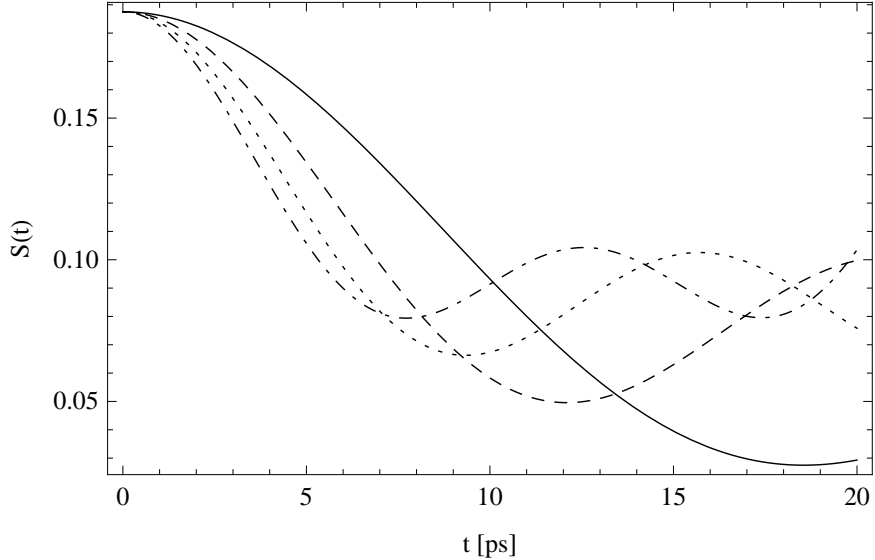}
\caption{The case of magnetization along the $z$-axis. $S(t)$ for $N$=1, 2, 3, 4 (lines: regular, dashed, dotted, dotdashed, respectively); n=10. }
\label{fig3r}
\end{figure}

The numerical calculations of~(\ref{S}) using~(\ref{expansion}) bring to the following statements:\\
1. Indeed, one can observe the decrease of $S(t)$ in the basis of p-Ps and o-Ps states (fig.~\ref{fig.1}), but nondiagonal terms do not disappear completely and permanently.
\\
2. After the decrease of $S(t)$ shown in figs~\ref{fig.1} and \ref{fig.2}
%
the oscillations were obtained\footnote{Oscillations are visible when calculated with sufficiently large $n$ to obtain reliable outcomes for $t > T_d$. Here maximally $n$=10 was applied which is enough to calculate reliably the evolution up to about 30~ps, so reliable oscillations are shown for the settings with enough small $T_d$. As an example see the calculation for $N$=4, $n$=10 shown in fig. \ref{fig.2} and in figs \ref{fig2r} and \ref{fig3r}} which are consistent with the results of calculations of decoherence for other complex systems in the spin bath \cite{Dobrovitski04,Dobrovitski04a}. For $N$=2, $T_d$ is\footnote{This result is not shown in any figure. Fig. \ref{fig.1} shows only a part of the whole $S(t)$ for this $N$ in order to show that each nondiagonal term constituting $S(t)$ collapses jointly but differently. The collapse of the whole $S(t)$ is shown in the next figures.} about 29~ps which means that the effect of the existence of the superposition of states lasts for the time detectable in PALS experiments and one should take into account this effect when decomposing the PALS spectra. This statement we consider as the most significant conclusion of this work.\\
3. Higher orders of hamiltonian in the evolution operator do not affect substantially the value of the decoherence time, within the meaning that $T_d$ may vary by few picoseconds, i.e. the decoherence maintains always the same time scale (fig.~\ref{fig.2}). Figure~\ref{fig.2} compares the calculations with the 3-rd and the 10-th power of $H$ as the highest order for a given $N$. Skipped parts of expansion of the evolution operator may change $T_d$ to some extent but their contribution is estimated in Appendix \ref{sec:appendix2}.
The aim of this paper is to estimate $T_d$ but not to calculate the time evolution of the system at any time. The calculation is performed with sufficient accuracy required to estimate the values of the $T_d$ for a given $N$. To calculate further evolution, next powers $n$ are needed. For example, in order to extinguish the increasing value of polynomials for the increasing time, $t>T_d$, one needs to calculate next powers of $H$ in the expansion of the evolution operator which give higher powers of polynomials dominating at large values of $t$; thus one obtains oscillations at times larger than $T_d$ instead of a rapid increase. Because the omitted terms of expansion may shift $S(t)$ for larger values of time, these terms may be more appropriate correction of $T_d$ for slow decoherence (small number of electrons in the bath giving larger $T_d$).\\
4. Decoherence time depends on a coupling constant $J$, which was set assuming that the interaction energy of Ps and the electrons from the wall is about 10$^{-4}$~eV (this is the energy which could convert ortho and para states and cause the initial superposition state in the bulk. Such a value of energy for the interaction of Ps in the free volume is reasonable only for some specific cases, e.g. when ortho-para conversion takes place. This process is characteristic only for specific substances \cite{Sharma88}). If the energy of the interaction is smaller, one can expect that the decoherence time increases (e.g. if the interaction energy is set to 10$^{-5}$~eV then $T_d\sim$150~ps for $N$=3 instead of $T_d$=18~ps as estimated above).\\
5. The value of $T_d$ decreases with the number $N$ of electrons in the environment reaching the asymptotic value 12.2~ps (fig.~\ref{fig.3}).
\\
6. If one assumes in the numerical calculations the magnetization of the medium (all spins in the same direction), like paramagnetics at extremally low temperatures, one obtains the suppression of the decoherence: it makes $T_d$ shorter for a given $N$ and the minimum of $S(t)$ becomes shallower, the shallower the greater is $N$.
Figure \ref{fig2r}
shows the comparison of an averaged $S(t)$ for random initial vectors of the state (lack of magnetization) and for all spins 'up' in the initial vector (presence of magnetization). In the case of magnetization the minimum is shallower\footnote{We decided to compare in this figure an averaged vector representing the state without magnetization with the state with magnetization where all spins of the environment are 'up'. The reason for averaging is
that particular states of the environment differentiate the state of the whole system to some considerable extent and the only way to compare the results with magnetization is to average them.}.
Furthermore, fig.~\ref{fig3r} shows $S(t)$ for different $N$ for magnetization. One can see that for greater $N$ one obtains shallower minima.\\
Calculations concerning magnetization were performed when all spins are situated along different direction from that we considered Ps state. The result is shown in fig.~\ref{fig2r}. A systematic study of $T_d$ dependence on the direction of electron spins was not performed.\\
The effect of inhibiting the decoherence in the presence of magnetization along the $z$-axis is for us a new one and we regard it as a hypothesis to be proved. The situation resembles the Ps eigenstates in the magnetic field~--~it is known that in the case of external magnetic fields the eigenstates of respective hamiltonian are those which are mixtures of p-Ps and o-Ps. Here supposing the magnetization of the sample one introduces a magnetic field. For this case the superposition of p-Ps and o-Ps seems to be maintained (which is indicated by the shallow minimum).
\\
The case of the magnetization presented above cannot be treated as one for the ferromagnetic sample. In ferromagnetic crystals Ps does not form. Such property of the bath may be rather connected with those of some molecules with the magnetic moment.\\
7. The Heisenberg hamiltonian leads to reasonable results in the case of positronium decoherence problem but it shows that the decoherence is not complete in the basis of p-Ps and o-Ps (there exists at any time the population of Ps atoms in the superposition state). The residue means that there exists at this time a Ps population with not well defined spin value or, in other words, not all Ps atoms are in the spin state. The probability to find such an undefined spin can be calculated as $1-(|\langle p-Ps|Ps(t)\rangle|^2+|\langle o-Ps|Ps(t)\rangle|^2)$ (it is neither p-Ps nor o-Ps). The smaller is $S(t)$ value at minimum, the more Ps have finally well defined spin.\\
We believe that the existence of the residue may be an artefact of the Heisenberg hamiltonian being the simplest way of treatment of the realistic case. We intend to follow the calculations with a more complicated hamiltonian, where $J$ is time dependent (Ps atom does approach to or goes away from the wall, so the interaction strength changes).\\
%
\section{Conclusions}
Our theoretical study of the decoherence of Ps in matter results in statement that cohered Ps transforms into the states with well defined spin and that the decoherence lasts enough time (picoseconds and more) to modify the positron lifetime spectra and momentum distributions and should be taken into account when any interpretation of the results in positron studies is performed. We have shown the way of modification for formulas used for PALS spectra decomposition.\\
The time of the decoherence decrease exponentially with the number of electrons interacting with Ps in the free volume of matter. We examined the influence of magnetization to be expected on the process of the decoherence.\\
Our calculations are consistent, according to the formalism, with the previous ones published for similar quantum systems and the results add the new aspect to the recent theory of Ps formation and its evolution in matter.
\appendix
\section{\label{sec:appendix1}}
The influence of hypothetical coherence of Ps on the PALS spectrum seems to be a complex problem which requires more study. As an example where this influence can be observed we propose the following one.\\
To interpret the PALS spectra one considers the formula for coincidence (2-~and 3-~quantum) rate ($N_{\gamma}$) resulting from the formula (see also \cite{He05})
\begin{equation}
dN_{\gamma}\sim-dP_e^{(\lambda_e)}-dS_o^{(\lambda_e)}-dS_p^{(\lambda_e)}-d(pPs)^{(\lambda_{pPs})}-d(oPs)^{(\lambda_{oPs})},
\label{coincCLASS}
\end{equation}
where: $P_e$- amount of free positrons, $S_p$, $S_o$- the same for pre-para-Ps and pre-ortho-Ps, $pPs$, $oPs$- amount of para- and ortho-Ps, respectively; $\lambda$s are the decay rates for a particular type of particles.\\
Furthermore, one assumes $P_e$, $S_p$, etc. obey the following equations
\begin{equation}
\begin{cases}
d P_e(t)= -(\lambda_e+\nu) P_e(t) dt, & P_e(0)=1,\\
d S_o(t)= \frac{3}{4} \nu P_e(t) dt -(\lambda_e+K) S_o(t) dt, & S_o(0)=0,\\
d S_p(t)= \frac{1}{4} \nu P_e(t) dt -(\lambda_e+K) S_p(t) dt, & S_p(0)=0,\\
d (oPs)(t)= K \  S_o(t) dt-\lambda_{oPs} (oPs)(t) dt, & (oPs)(0)=0,\\
d (pPs)(t)= K \  S_p(t) dt-\lambda_{pPs} (pPs)(t) dt, & (pPs)(0)=0,
\end{cases}
\label{dynamCLASS}
\end{equation}
where $\nu$, $\lambda_e$, etc. are the constants describing creation and decay rates of each population.
\\
If one considers an annihilation in matter where the pick-off process dominates, (\ref{coincCLASS}) relates mainly to the 2-quantum process of annihilation and both (\ref{coincCLASS}) and (\ref{dynamCLASS}) allow to calculate the intensity of the annihilation rate for the set of particular channels (i.e. for $P_e$, $(oPs)$, etc.). The decomposition of PALS spectra is a direct consequence of the use of formulas above.\\
If one supposes the existence of the coherence of the states, one does not distinguish the spin value for pre-Ps and one assumes the existence of Ps (and also pre-Ps) in the superposition state, which annihilates with the rate $\lambda_M$. So, one can replace (\ref{dynamCLASS}) and (\ref{coincCLASS}) by
\begin{equation}
\begin{cases}
d P_e(t)= -(\lambda_e+\nu) P_e(t) dt, & P_e(0)=1,\\
d S(t)= \nu P_e(t) dt -(\lambda_e+K) S(t) dt, & S(0)=0,\\
d (oPs)(t)= K \  S(t) dt-\lambda_{oPs} (oPs)(t) dt, & (oPs)(0)=0,\\
d (pPs)(t)= K \  S(t) dt-\lambda_{pPs} (pPs)(t) dt, & (pPs)(0)=0
\end{cases}
\label{dynamNEW}
\end{equation}
and
\begin{equation}
dN_{\gamma}\sim-dP_e^{(\lambda_e)}-dS^{(\lambda_e)}-d(MS)^{(\lambda_{MS})}-d(pPs)^{(\lambda_{pPs})}-d(oPs)^{(\lambda_{oPs})}-d(MPs)^{(\lambda_{MPs})}.
\label{coincNEW}
\end{equation}
The formula (\ref{coincNEW}) is not a simple replacement of (\ref{coincCLASS}) because the last term in it describes the decays where the number of photons per one annihilation act is not fixed and, for example, 3 quantum decay cannot be neglected here.
\\
\section{\label{sec:appendix2}}
\begin{figure}
\centering
\includegraphics[width=0.45 \textwidth]{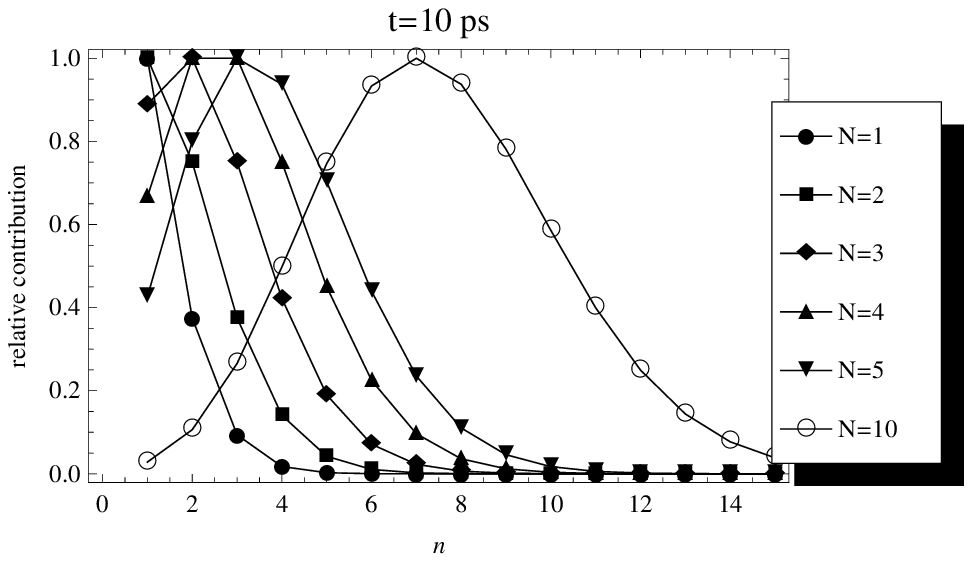}%
\hspace{0.1 \textwidth}%
\includegraphics[width=0.45 \textwidth]{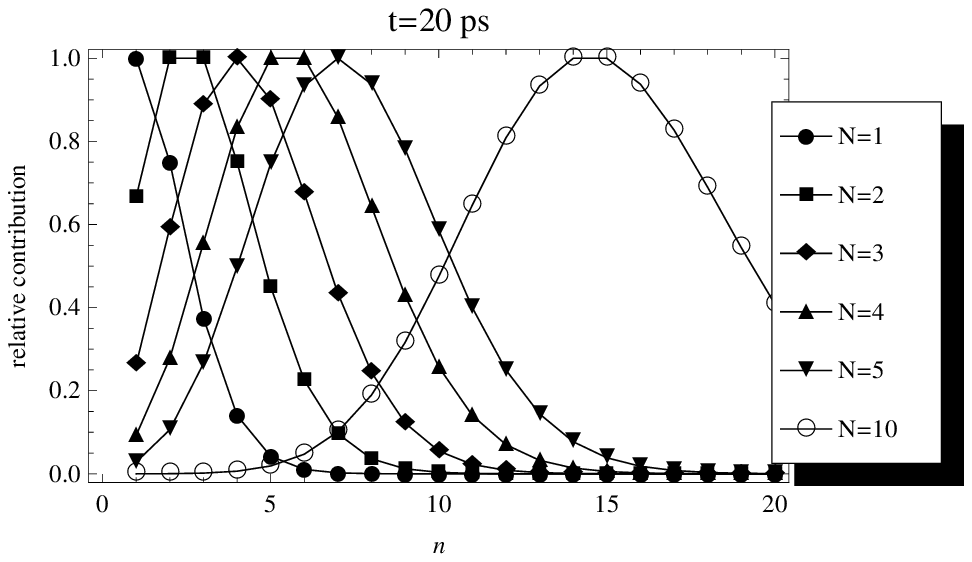}
\caption{The boundary value of the absolute of $n$-th term (normalized to the maximal value) in (\ref{expansion}) showing its contribution at $t$=10~ps and $t$=20~ps to approximate the exact result in the region of calculated $T_d$.}
\label{szacowanie}
\end{figure}

The $H^n$ operator can be written as
\begin{equation}
H^n=(\sum_{p=3}^{N+2}\sigma^{(2)}\circ \sigma^{(p)})^n=\sum_{p,s,t,\dots}^{n\text{-fold sum}}\sigma^{(2)}\circ \sigma^{(p)}\cdot \sigma^{(2)}\circ \sigma^{(s)}\cdot\dots.
\label{rozwiniecieHdoN}
\end{equation}
Each superscript changes in the range $(3,N+2)$, so we have $N^n$ different terms with the fixed value of indices. Because $\sigma^{(2)}\circ \sigma^{(p)}=\sum_{i=1}^3\sigma_i^{(2)}\cdot \sigma_i^{(p)}$ we have $3^n$ different values of each term summed in (\ref{rozwiniecieHdoN}) (for fixed $p$, $s$, $t$, $\dots$). Each of such terms can give maximally the value $1$ on any state\footnote{We assumed a very coarse approximation: if one of $\sigma_i^{(p)}$ gave 1 in the extreme case, then none of $\sigma_j^{(p)}$ for $i\not= j$ could reach this value. In spite of this such extreme value is applied for simplicity. Because of this simplification the approximation is overestimated.} so the resultant coefficient related to the $n$-th power of hamiltonian in (\ref{expansion}) is $\aleph/n!\cdot (-iJ\hbar/4\cdot t)^n$, where $\aleph=(3\ N)^n$. Substituting the assumed value of $J$ into this coefficient we can see the contribution of the $n$-th term as a function of $t$. This contribution for fixed $t$=10~ps and $t$=20~ps (the region of $T_d$ for several $N$) is shown in figure \ref{szacowanie}.
\\
If one knows the contribution of skipped terms in the evolution operator at any $t$ one can estimate the expected shift of minimum ($T_d$) of polynomial we obtained from truncation at smaller $n$ and that one expected when more accurate calculations could be performed. For doing so, see Appendix \ref{sec:appendix3}.
\section{\label{sec:appendix3}}
Let us suppose we have two polynomials $f_1(t)=\sum_{i=0}^{k}a_i\ t^i$ and $f_2(t)=\sum_{i=0}^{k+c}b_i\ t^i$ which represent the $S(t)$ function calculated when $n$=k and $n$=k+c respectively. Let single out $s$-th minimum of each polynomial, $t_{1s}$, $t_{2s}$, and relate it to those constituting $T_d$. Suppose we know (see Appendix \ref{sec:appendix2}) amount of correction at any $t$ given by $c$ additional terms of expansion of evolution operator; denote this surplus as $\delta(t)$. This function $\delta(t)$ can be expressed as
\begin{equation}
\delta(t)=f_2(t)-f_1(t)=\sum_{i=0}^{k}(b_i-a_i) t^i+\sum_{i=k+1}^{k+c}b_i t^i.
\label{gwiazdka}
\end{equation}
Let us construct the auxiliary function
\begin{equation}
F=f_2(t)-f_1(t-\Delta t),
\label{posilkowa}
\end{equation}
which has, if $t$=$t_{2s}$, a maximum for $\Delta t$=$t_{2s}-t_{1s}$. Inserting the definitions of $f_i(t)$ into (\ref{posilkowa}) and using the Newton formula for the binomial expansion we have
\begin{equation}
F=\delta(t)+\sum_{i=0}^k[i\ a_i\ t^{i-1}\ \Delta t-a_i\frac{i(i-1)}{2!}t^{i-2}(\Delta t)^2+\cdots]
\label{posilkowaInsert}
\end{equation}
and finally inserting into (\ref{posilkowaInsert}) the value $t_{2s}$ we find a local maximum of $F$ as
\begin{equation}
\delta(t_{2s})+\sum_{i=0}^{k}[i\ a_i\ t_{2s}^{i-1}\Delta t-a_i\ \frac{i(i-1)}{2!}t_{2s}^{i-2}(\Delta t)^2+\cdots]=max.
\end{equation}
This equation is equivalent to the derivative of its left side over $\Delta t$ equated to 0. This equation should give a maximum value for $\Delta t=t_{2s}-t_{1s}$, i.e.:
\begin{equation}
\frac{\partial F}{\partial \Delta t}|_{\{t=t_{2s},\ \Delta t=t_{2s}-t_{1s}\}}=0.
\end{equation}
The polynomial $f_1(t)$ is usually known as it is the solution of our problem for truncation of the evolution operator at smaller $n$. For known values of $t_{1s}$ and $a_i$ one obtains an algebraic equation for $t_{2s}$. In this way we can calculate the expected difference between known $T_d$, obtained for shorter expansion of the evolution operator, and that one if more accurate calculations could be performed.
\begin{acknowledgments}
The authors want to thank Dr. T.~Paterek (from the Centre for Quant. Technol. of National University of Singapore) for the introductory discussions and Dr. M.~Turek (Inst. of Phys., M.~Curie-Sk{\l}odowska University, Poland) for the encouraging word. We thank also M.~Opala (computer farm administrator at Dept. of Theoretical Physics, UMCS) for help in adjustment of our work on the accessible computer hardware.
\end{acknowledgments}

\end{document}